\documentstyle [12pt]{article}
\evensidemargin\oddsidemargin
\begin{document}
\count0 = 1
\begin{titlepage}
\vspace{30mm}
 \title{{ Relational Quantum Measurements,\\    
 Information, and State Collapse  \\ }}
\author{S.N.Mayburov 
 \thanks{E-mail ~~ maybur@sgi.lpi.msk.su  ~~
 }\\
Lebedev Inst. of Physics\\
Leninsky Prospect 53\\
Moscow, Russia, 117924\\
\\}
\maketitle
\begin{abstract}

The  quantum measurement problem considered
for measuring system (MS) 
 consist of measured state S (particle),
 detector D and information processing device (observer) O.  
It's shown that  O states selfreference structure results in 
principal nonobservability of 
MS  interference terms which discriminate pure and mixed
S states. Such  observables restriction permit to construct for MS states 
 subjective  representation (SR) which 
 describes probabilistic  evolution for 
measurement events observed by $O$ and his subjective  information
about S values.  SR is dual and  nonequivalent to 
MS Hilbert space $H$ for external observer $O'$. 
Due to it SR evolution is compatible
with Schrodinger linear MS  evolution observed by $O'$. 
It's  argued that SR  evolution    corresponds to S state
 collapse for individual events observed by $O$.
 
\end{abstract}
\vspace{20mm}
\small{Talk given on 'Quantum communications and measurements' conference\\
Capri , Italy, July 2000\\}
\vspace {20mm}
\vspace{20mm}
\end{titlepage}
\section { Introduction}
Despite that Quantum Mechanics (QM) describe perfectly most of
experimental effects in microscopic domain there still some
difficult questions and 'dark spots' connected with its Measurement
Theory  and more generally with its proper interpretation.
 Of them the problem of  the state vector collapse 
 is most remarkable and straightforward and it's
still open despite the multitude of the proposed
models and theories ( for the review see $\cite  {Busch}$). Eventually
the measurements and collapse studies can help us to
 select the true QM interpretation out of many proposed.
 This paper analyses some microscopic dynamical models of quantum
measurements which attempt to describe
  the  evolution of the measuring system (MS) from
the first QM principles. In our approach  MS includes
the measured state (particle) S,
  detector D  amplifying S signal, environment E  and
 observer $O$ which processes and stores the information. Under 
observer we mean information gaining and utilizing system 
(IGUS) of arbitrary structure  $\cite {Gui}$.
 It can be  human brain or some automatic device 
processing the information , but in both cases it's
the system with many internal degrees of freedom (DF) 
memorizing the large amount of information.
In general the computer information processing or perception  by human brain
 is the physical objects evolution which on microscopic level
supposedly obeys to  QM laws   $\cite {Alb}$.

 Copenhagen QM interpretation divide our physical world
into microscopic objects which obeys to QM laws and macroscopic objects
, also observers which are strictly classical. This artificial partition
was much criticized, first of all because it's not clear where
to put this quantum/classical border. Moreover there are strong
experimental evidences that at the dynamical level no such border
exists and QM  successfully describes large, complicated systems
including biological ones.    

The possible role of observer  
in quantum measurements was discussed for long time \cite {Wig}, but 
now it attracts the significant attention again  due to the
the progress of quantum information studies $\cite {Pen}$. The 
 different aspects of  observer inclusion in QM formalism 
was discussed  in Rovelli paper which includes extensive
 review of previous activity 
 $\cite  {Rov}$. Such approach 
 called Relational QM isn't selfconsistent formalism at the current stage.
 Rather  it's phenomenological theory with many $ad$ $ hoc$ assumptions
 especially concerning the measurement problem and
 we'll investigate some of its difficulties. As its particular 
 version 
 can be regarded Kochen Witnessing QM interpretation \cite {Koch}.

 Relational QM  concedes
(Hypothesis 1 of Rovelli paper) that QM description is applicable
both for microscopic  states  and macroscopic
objects including observer $O$ which  Dirack state vector
$|O\rangle$  can be defined relative to  some other observer $O'$, 
which is also another quantum object.
Of course this assumption it's not well founded and real $O$ state
can be much more complicated, but it's reasonable to start from
 that simple case.
 Consequently the evolution of any complex system C
 described by Schrodinger equation 
  and for   C
 the superposition principle holds true 
at any time. MS measurement description by $O$ 
formally  must include evolution of 
 $O$  own internal DFs  which  participate in the interaction with S 
 $\cite {Alb}$.

The role of observer in the measurement and its 
 selfmeasurement
  analyzed in some works as the implication
of more general algebraic problem of selfreference $\cite {Busch}$.
Following this approach Breuer derived the general selfmeasurement 
restrictions  for classical and quantum measurements  $\cite {Bre}$.
This formalism don't resolve 
the measurement problem,  but its comparison with
Relational QM  will be shown to result in some important 
conclusions. From this analysis we propose modification of standard
QM Hilbert space formalism  which  permit  $O$ to observe
 state collapse without  contradiction with MS linear evolution.
Its  main feature is the extension of QM states manifold permitting
to account observer selfmeasurement effects, which were qualitatively
formulated by Rovelli $\cite {Rov}$.  

In chap 2. we describe our measurement model 
and propose the particular variant of QM formalism modification.
In chap 3. we'll discuss $gedankenexperiments$ which
help to interpret our formalism and discuss its implications.
In chap. 4 we'll discuss 
 interpretation of our results and their 
physical and philosophical implications.

Of course if some correction to quantum dynamics like in GRW model
exist $\cite {Gir}$ then the state collapse can occur in macroscopic
detector. But until such effects would be found and the 
standard QM Hamiltonian  regarded well established,
 we must inspect
in detail  observer properties exploring Measurement problem.

Here it's necessary to make some technical comments on our model premises
and review some terminology.
For our  models we'll suppose that MS  always can
 be described completely (including Environment if necessary)
 by some state vector $|MS\rangle$ relative to 
$O'$ or by density matrix if it's in the mixed state.
 MS can be closed system , like atom in the box or open pure
system  surrounded by electromagnetic vacuum or  E of other kind.
We don't assume in our work any special properties of $O$ internal states
beyond standard QM. 

 We'll use Rovelli approach to
 quantum information as the correlation between  S and $O$ states.
It differs from the standard definition $\cite {Shum}$,
 but is more useful for pure states measurements.
It defines that $O$ have the information about S, if $O$ (internal) 
 state correlates with S state $\cite {Rov}$. Note that such correlation,
 if to exclude the noice and errors
 means  S and $O$ states causal connection realized via their
interaction.
 In particular we'll  use projective or selected
information $I_Q$ related to arbitrary  S observable $Q$.
   $I_Q$ measure can be 
some  function of Q dispersion of the kind $I_Q\approx \sigma_Q^{-1}$.
 More correct seems to use $I_Q=-ln\sigma_Q^2$ with proper normalization,
 but our conclusions only slightly depend on its exact form
$\cite {Busch}$.
$\sigma_Q$ derived from $O$ state  after the measurement of some 
   $Q'$ on S, which in general can differ from $Q$ but gives
some Q value estimate.

  In this paper  the brain-computer analogy used without
discussing its reliability and philosophical 
implications $\cite {Pen}$. We'll ignore here quantum computer options
 having in mind only the standard solid-state dissipative computers.
We must stress that throughout our paper the observer consciousness
 never referred directly.
 Rather in our model observer can be regarded as active 
reference frame (RF) which interacts with studied object.
 The terms 'perceptions', 'impressions' used by us
 in a Wigner sense $\cite {Wig}$ of observer subjective description
of experimental  results 
and so  can be defined in strictly physical and information
theory terms.

\section {Selfmeasurement and Weak Collapse}
 
We'll consider the measurement description 
for simple model for detector D and observer $O$ each only with one  DF 
each corresponding to Von Neuman scheme.
The example of dynamical model with many DFs
gives Coleman-Hepp model described in appendix $\cite {Hep}$.
Account of many DFs doesn't change principally the results
obtained below \cite {May3}, but
in addition it resolves the problem of 'preferred basis'
arising for one  DF detector model $\cite {Elb}$.  
 Let's consider in this ansatz  $O'$ description of
 the measurement by $O$ of binary observable $\hat{Q}$ on S state : 
 $$
\psi_s=a_1|s_1\rangle+a_2|s_2\rangle
$$
, where $|s_{1,2}\rangle$ are $Q$ eigenstates with eigenvalues $q_{1,2}$.
Initial D, $O$ states are $| D_0\rangle, |O_0\rangle$
relative to $O'$  RF. 
We assume that  S-D-$O$ measuring interaction starts at $t=t_0$
and finished at some finite $t_1$.
 It follows from the linearity of Schrodinger equation
that for suitable interaction Hamiltonian $\hat{H}_I$
 at $t>t_1$ 
 the state of MS system  relative to $O'$ observer is
\begin {equation}
   \Psi_{MS}=a_1|s^f_1\rangle|D_1\rangle|O_1\rangle+
a_2|s^f_2\rangle|D_2\rangle|O_2\rangle
                                   \label {AA1}
\end {equation}
Here $|D_1\rangle, |O_{1}\rangle$ are D,$O$ state vectors
obtained after  the measurement of particular $Q$
eigenstate  $|s\rangle=|s_1\rangle$  and are eigenstates
of $Q_D, Q_O$ observables and  correspondingly for $s_2, O_2$ ( below state
 vectors with $n>2$ components used with the same notations).
 All this states including $O$ belongs to Hilbert space
$H'$ defined in $O'$ RF and Hilbert space $H$ in $O$ RF supposedly
can be   obtained performing unitary $H'$ transformation $\hat{U}'$
 to $O$ c.m.s.. $\hat{U}'$  can be neglected if only internal
or RF independent discrete states regarded permitting to take $H=H'$.
 For realistic IGUS $|O_{1,2}\rangle$ can correspond to some
excitations of $O$ internal collective DF
 like phonons, etc., which memorize this $Q$ information,
 but we don't consider its particular physical
mechanism here.

 For the  simplicity in the following  we'll  omit detector D in 
 MS chain
assuming that S directly interacts with $O$. It's reasonable for our 
simple model,
because if to neglect decoherence the only  D effect is
the amplification of S signal to make it conceivable for O.
We'll assume also that after interaction S leaves $O$ volume, which
can be regarded as 'self-decoherence', because final MS state quantum
phase becomes unavailable for $O$. More realistic dynamic decoherence
mechanisms will be discussed in final chapter $\cite {May3}$. 
In this case MS initial state is :
\begin {equation} 
     \Psi^0_{MS}=(a_1|s_1\rangle+a_2|s_2\rangle)|O_o\rangle  \label {AAB}
\end {equation}
and final state at $t>t_1$ :
\begin {equation}
   \Psi_{MS}= a_1 |s^f_1\rangle|O_1\rangle+
a_2|s^f_2\rangle|O_2\rangle
                                   \label {AA2}
\end {equation} 
to which corresponds the density matrix ${\rho}_{MS}$.
In most cases one can take for the simplicity $|s^f_i\rangle=|s_i\rangle$
without influencing main results.
We'll suppose that for $t>t_1$ measurement definitely finished which
simplifies all the calculations
, but in fact that's fulfilled exactly only for the restricted
class of models like Coleman-Hepp. 
Thus QM predicts  at time $t>t_1$  for external observer $O'$ 
MS is  in the pure state  $\Psi_{MS}$ of (\ref {AA2}) which is superposition of
 two states.
Yet we know from the experiment that 
   $O$ observes some definite random $Q_O$ value
and acquires the  state $O_1$ or $O_2$, from which he concludes
that  S state is $|s_1\rangle$ or $|s_2\rangle$.
If detector D included into MS chain
then this $O$ memory states results from observing
 detector pointer position $D_1$ or $D_2$, but if D omitted  like in our
case they
result from direct S-$O$ interaction.

The standard QM  conclusion  is that MS final  state coincides with
the statistical ensemble of such individual final states for $O$
described by density matrix 
 of mixed state $\rho_m$ which  presumably
means the state collapse:
\begin {equation}
 \rho_m=  |a_1|^2|s_1 \rangle \langle s_1||O_1 \rangle \langle O_1|+
|a_2|^2|s_2 \rangle \langle s_2||O_2\rangle \langle O_2|
                                                              \label {AA3}
\end {equation}
It would be natural to expect that MS  final states described by $O$ 
and $O'$ are connected by some unitary transformation, but
it's well known that such transformation between (\ref {AA3})
and (\ref{AA2}) don't exist $\cite {Busch}$.
It's quite difficult to doubt both in correctness of
 $O'$ description of MS evolution
by Schrodinger equation and in collapse experimental observations.
 This  contradiction constitutes famous Wigner dilemma
for $O, O'$  $\cite {Wig}$. 
Really if for $O$ MS state is (\ref {AA2}) it formally
describes the superposition of alternative $O$ impressions
on measurement results and its meaning is difficult to interpret.
 We mentioned already that Relational QM
 suppose phenomenologically that $O'$ description
 is correct. But in standard QM formalism it's incompatible with 
assignment of definite values $q_i$ to
Q in $O$ RF  in particular event \cite {Lah}.
We'll propose here alternative formalism in which
 MS state linear evolution description by $O'$ is correct, but the
description (\ref {AA3}) for $O$ is incomplete.

Because we include observer in our model it's necessary to formulate
some minimal assumptions about observer internal states  which
stipulate his  reaction  on the input quantum signal.
We'll suppose that for any Q eigenstate $|s_i\rangle$ 
with probability close to 1 after S measurement finished at $t>t_1$ 
and $O$ state becomes $|O_i\rangle$ observer $O$ have the impression
that  the measurement event occurred and the value of
outcome is $q_i$.  It means
that at least in this case and also for mixed states the signal
memorization described by Schrodinger equation and so it is reversible
process which for $O$ percepted as individual event with definite outcome.
 Of course this is only subjective
not universally objective event, but probably other kinds of events are 
impossible.
This events or impressions are 
connected with the excitations of $O$ internal DFs. The simple $O$
toy-model of information memorization is hydrogen-like atom
 for which $O_0$ is ground state
and $O_i$ are different metastable  levels excited by $s_i$.  
  If S state is the superposition (\ref {AA1}) then we'll
suppose that its measurement also  result in appearance of some new $O$ 
impression without specifying it at the moment.
This considerations have little importance for the following calculations,
rather they explain our philosophy of impressions-states relation.

Here we assumed that at $t>t_1$ measurement finished with probability
$1$, but for realistic measurement Hamiltonians
 it's only approximately true, because  transition amplitudes have 
long tails. This assumption exactly fulfilled only for some simple
models like Coleman-Hepp, but we'll apply it  to simplify our analysis.   
The more subtle question of exact time at which $O$ percepts its
own final $O_j$ isn't important at this stage.
 
To discuss $O$ selfdescription  let's consider Breuer selfmeasurement
theorem which is valid for the large class  both of
classical and quantum measurements $\cite {Bre}$.
Any measurement of studied system MS is the 
mapping of MS states set $N_S$ on  observer states set $N_O$.  
For the situations when observer $O$ is the part of the studied system MS 
- measurement from inside, $N_O$ is $N_S$ subset and 
$O$ state in this case is MS state projection on $N_O$
 called MS restricted $O$ state $R_O$.  
From $N_{S}$ mapping properties some principal restrictions for
$O$ states were obtained. The theorem claims that if for two arbitrary MS 
states their restricted  $O$ states coincide then for $O$ this MS 
states are indistinguishable.
 The origin of this effect is easy to 
understand : $O$ has less number of DFs then MS and so can't describe
completely MS state.
For quantum measurements Breuer supposed  that O restricted state
can be the partial trace of complete MS state  (\ref {AA2}) :
\begin {equation} 
   R_O=Tr_s  {\rho}_{MS}=\sum |a_i|^2|O_i\rangle\langle O_i|
      \label {AA4}
\end {equation}
Note that  for MS mixed state $\rho_{m}$
of (\ref {AA3}) the restricted state is the same $R^m_O=R_O$.
This equality doesn't mean collapse of MS state $\Psi_{MS}$, because
it holds for statistical quantum ensembles, but collapse in fact must
 be tested  for individual events as explained below. 
Such restricted $R_O$ form suppose
 that O can percept only his internal
excitations independently of quantum correlations with  S state.
This assumption can be wrong for quantum
systems due to well known quantum entanglement.
and in fact  this effects study shows  that from equality of 
restricted states doesn't follows the transition of pure system 
 state to mixed
one.  MS interference term observable :
\begin {equation}
   B=|O_1\rangle \langle O_2||s_1\rangle \langle s_2|+j.c.
    \label {AA5}
\end {equation}
being measured by $O'$ gives $\bar{B}=0$ for mixed MS state (\ref {AA3})
and in general $\bar{B}\neq 0$  for pure MS state (\ref{AA2}).
It  evidence that even for statistical ensemble 
the observed by $O'$ effects can differentiate  pure and mixed MS states.
Note that $B$ value principally can't be measured by $O$ directly, because
$O$ performs $Q$ measurement and $[Q,B] \neq0$ $\cite {May3}$. 

Considering individual events Breuer notices that for mixed incoming state S
their MS state is :
$$
\rho^m_l=|a_l|^2|O_l\rangle \langle O_l|| s_l\rangle\langle s_l|
$$
with arbitrary $l$ which in accordance with standard QM objectively
exists, but can be unknown for $O$.
Its restricted state $R^m_l=|a_l|^2|O_l\rangle \langle O_l|$
 and so differs from 
$R_O$. Due to it main condition of Breuer  Theorem violated
and so it don't applicable for this problem.
 From that  Breuer concludes that the restricted states
ansatz doesn't prove the collapse appearance 
  for individual events even with inclusion of observer in 
standard QM formalism with Hilbert space states set. 
 The analogous conclusions follows from the critical
 analysis of Witnessing interpretation $\cite {Lah}$.

  Breuer analyses is quite informative and useful
 for our attempt to modify QM formalism, because  
it prompts the particular extension of MS states set $N_S$ 
which can describe MS state collapse noncontroversially.
We'll demand that this extension suits to
 Relational QM hypothesis  that $O$ and $O'$ can make different
conclusions about MS measurement - MS  final states relative to
$O$ and $O'$ are nonequivalent.
Due to it results of $B$ measurement by $O'$ can be unimportant
for $O$ description of measurement and so at least statistically
it can be the same for pure and mixed MS state.
For individual events to agree with Relational QM our formalism
 is demanded to describe the weak (subjective)
collapse  having different conditions then standard one.
It means the following : For MS final state its description  (perception)
by observer $O$ presents
the probabilistic events realization for $O$ with partial  probabilities
$|a_i|^2$. It means that any restricted statistical state $R_O$ has
unique physical realization for $O$ which coincide for pure and mixed
S state with the same  probabilities $|a_i|^2$.
 This is subjective collapse observed only by $O$
and in the same time MS state for $O'$ stays pure and evolves according
to Schrodinger equation.

So our  aim is to find minimal
 modification of QM states set  
which can incorporate simultaneously both MS linear state
evolution for $O'$ and random events observed by $O$.
$B$ nonobservability for $O$  hints that states manifold
 must be modified for  MS states description, because in standard 
Hilbert space all hermitian operators are observables.
 It's worth to remind that Hilbert space
is in fact empirical construction which choice advocated by fitting
most of QM data, and so QM states set modification doesn't seems
 unthinkable in principle.
 Such modifications attempts were 
published already and most famous is Namiki-Pascazio many Hilbert spaces
 formalism.
In standard QM formalism all its states manifold representations
are unitarily equivalent, but observers interactions and
evolution aren't considered in it.
It will be shown that new formalism
  to some extent is analog of the
nonequivlent representations of commutation relations. Analogous
superselection systems
are well studied for quantum nonperturbative Field theory (QFT)
 with infinite DF number
$\cite {Ume}$. QFT methods were applied to measurement problem,
 but it's not clear its applicability for finite even macroscopic
systems $\cite {Fuk,May2}$. 

To illustrate the formalism features let's regard how
 $O$ measures some stochastic parameter $q$ with distribution $P(q)$
 in Classical Physics.
For some $q$ distribution $P(q)$ (or $P_i$  array for discrete $q$ )
when $O$ measures $q$ he acquires instantly information about $q$ value
and initial $O_0$ state changes to some $O_j$ correlated with
measured $q_j$. Formally
 at this moment $P(q)$ collapses to delta-function, but
    in classical case
 this effect reflects only $O$ information change.
Suppose that $q$ acquires discrete random values with 1- dimensional
 probabilities matrix $P_l, (l=1,n)$
and after its measurement $O$ acquires  state $O_i$ with probability $P'_i$
which in ideal case coincides with $P_i$ and define $O$ ensemble statistics.
To present  recorded result $O_i$ in given event 
 random 1-dimensional matrix  $V^O=(0,...,O_j,...,0)$ is used
 with only one $v^O_j \neq 0$ in  each event. We  write formally this
 event-state manifold 
 as $N_{cl}=P \bigotimes V^O$  assuming to apply 
 in quantum case such dual form which unite probability
distribution $P$ and $O$ information in event $V^O$.  

It's worth to  remind that  experimentalist
never observes state vector directly, but his data consists of
individual random events like detector pointer counts and 
  the initial state vector restored  from
observed random events statistics. This fact prompts us to explore
 QM dual representations, in which state vector and random events
can coexist simultaneously.
The phenomenological dual ansatz for statistical ensemble states
was proposed in $\cite {May3}$.  
 We describe first this
 new representation  for our MS system evolution observed
by $O$ and $O'$ which Hilbert spaces $H, H'$ will be our starting point.
We use QM density matrices manifold $L_q= ( \rho \geq 0, Tr \rho=1)$
constructed of $H$ state vectors. They evolve according to standard
Schrodiner-Liouville equation  with MS Hamiltonian $\hat{H}_c$ :
\begin {equation}
    \dot\rho=[\rho,\hat {H}_c]  \label {AA8}
\end {equation}

 For initial MS state vector $\Psi^0_{MS}$
  MS measurement result in $\rho_{MS}$ of (\ref {AA2}).
 Inside $L_q$ we extract $O$ restricted
states submanifold of (\ref {AA4}) $R_O=Tr_s \rho$ and calculate in $O$ basis
the weights matrix $P_j(t)=Tr_O (\hat{P}^O_j R_O)$ where $\hat{P}^O_j$
is $O_j$ projection operator. For (\ref {AA2}) it  gives $P_j(t)=|a_j(t)|^2$.
 As we stressed already discussing Breuer Theorem the
restricted state $R_O$ of (\ref {AA4}) by itself has no probabilistic
 meaning and $Q$ values are uncertain and nonobjective
 in $O'$ RF $\cite {Bre}$. But
in $O$ RF in our dual framework we'll suppose it becomes
 probabilistic distribution $P_j$ which generates random 
 $O$ states
 $V^O=  |O_j\rangle \langle O_j|$ for given event.
It describes the restricted subjective state
 $|O_j\rangle \langle O_j|$ observed
by $O$  in given event and corresponding to random $q_j, q_{Oj}$ values.
But in distinction from standard QM in our
formalism  the complete  state  $\rho$ don't disappear
after the measurement. On the opposite 
it evolves all the time according to Schrodinger-Liouville equation
 (\ref{AA8}) and  doesn't suffers the collapse stochastic jumps.
Thus our dual state or event-state, which due to $j$ randomness
 differs for each event
 in $O$ RF is doublet $\Phi=|\rho,\, |O_j\rangle \langle O_j| )$.
Thus it includes dynamical (objective) component $\rho$ which evolves linearly
according to (\ref {AA8}), but isn't observed directly by $O$ 
 and  $O$ subjective component $V^O$ which describe $O$
impression about event and which probabilistic
evolution controlled by $\rho$.
In this case
 $O_j$   observes with probability $1$  S state
component  with which it's entangled i.e. $s_j$ and MS  subjective state
component  can be defined  :
\begin {equation}
  V^{MS}=  |O_j\rangle \langle O_j| |s_j\rangle \langle s_j| \label {AA99}
\end {equation}
, but  for our problem it's equivalent to $V^O$.

Complete manifold in $O$ RF for this event-states is
$N_T=L_q \bigotimes L_ V$ i.e  tensor product of dynamical and
subjective components. $L_V$ is the  linear space
of diagonal positive matrices with $tr V^O=1$.
 If we restrict our consideration only to
pure states then $N_T$ is equivalent  to $H \bigotimes L_V$
 and  state vector can be used as event-state dynamical component.
Note that the physical meaning of
Hilbert space $H$ in our formalism  essentially differs from standard
QM, because the  operator $B$ of (\ref {AA5}) isn't observable for $O$.
 Before measurement starts
  event state is
 $\Phi_0=|\Psi^0_{MS},\, V_0^O ) $ where $V_0^O=|O_0\rangle \langle O_0|$
describe $O$ definite initial information.

In our formalism
$O'$ has its own  subjective linear space $L'_V$  
 and in his RF  the event-states
 manifold is $N'_T=H' \bigotimes L'_V$ for pure states.
 From the above description it seems that subspace
$L_V$ must be unobservable for $O'$ and vice versa, because
$O'$ interacts directly only with $\rho$ component of  event-state.
But this is true if $O', O$ don't interact  and
 $V^O, V'^O$ events can be correlated if $O'$ measures $O$ state
 expressed by $\rho$
and  below we'll discuss such effects.

 If S don't interact with $O$ then $V_0^O$ is time invariant and one obtains
standard QM evolution for event-state dynamical component $\rho$.
If one interested only to calculate $\bar Q $ after S measurement by $O$ 
or any other expectation values ignoring event structure
it's possible to drop $V^O$ component and to make standard QM calculations
for $\rho$.

Proposed doublet states  ansatz can be regarded as the upgrade of standard
reduction postulate, which  describes for standard QM how state vector
correlates with the changes of observer information in the measurement.
The main difference is  that in place of abrupt and irreversible
 state vector $\Psi_0$ reduction to some random state vector $\Psi_l$ in 
standard QM, in our formalism in  $O'$ RF the dynamical component $\Psi_{MS}$
of  MS event-state evolves linearly and reversibly
in accordance with (\ref {AA8}). 
It's only subjective component $V^O$ which changes abruptly and 
probabilistically describing the change of
$O$ subjective information about $S$.
We must stress that 
subjective  $V^O$ component is physical object
which is new degree of freedom  connected with $O$ internal
state, which lays outside of MS Hilbert space $H$.
 We noticed already that in standard QM formalism in MS state (\ref {AA2})
Q values aren't objectively existing for $O'$ and $O$
 which is serious argument against
Witnessing interpretation $\cite {Lah}$. That's the same for $O'$
  in our formalism, but  in the same time $Q$ and $Q_O$ can have
objective values $q_j, q_{Oj}$ in $O$ RF.

 Equation (\ref {AA8}) is in fact master equation for
probabilities $P_j(t)$ which induce $V^O$ probabilistic  distribution
which becomes new random DF of final $O$ states $O_j$.
Due to independence of MS dynamical state component
of internal parameter $j$ of $V^O$ 
this $O$-S evolution is reversible. Due to it no experiment performed
by $O'$ on MS wouldn't contradict to standard QM. If $O$ perform
selfmeasurement experiment on MS the situation is more subtle and
 will be discussed in the next chapter.  Note that in this formalism
parameter $j$ don't existed before S-$O$ interaction starts. 

Now we regard in more detail relation between dynamical and subjective
components of event-state. We proposed already that parameter $j$ of $V^O$
defined at random with probabilities  $P_j$ in  S measurement. In general
to calculate  $\Phi$ evolution for arbitrary complex system MS
Schrodinger-Lioville equation
for  MS Hamiltonian can be used and $P_j(t)$ found.
 Then from $P_j(t)$ at any time 
we find random $V^O$. So if we have several S-$O$ rescatterings
each time after it we get new $V^O$ state component
which effect in details will be discussed below.
 To exclude spontaneous $V^O$ jumps
without effective interactions with external world 
 we introduce additional $O$ identity condition \cite {Sna} :
 if S and O don't interact   then
 the same   $V^O$ conserved.
We supplement it by more general condition : if  different $\Psi_{MS}$
$O_j$ branches don't intersects i.e.
 $\langle O_i|\hat{H}_c|O_j\rangle=0 , i \neq j$ then $V^O$ conserved.  
It means that $O$ observes 
 constantly only $s_j $ branch of S state, despite that after
measurement $s_j$ state can evolve.
This conditions don't influence MS dynamics, but only  subjective 
 information $V^O$.
Here $L_V$ corresponds to the simplest measurement and in general it  
will can have more complicated form, which can be obtained 
demanding them to correspond to
the probabilities definition of standard QM $\cite {Busch}$. For
example 2-dimensional values correlation measurement by $O$
 has the distribution :
\begin {equation}
P_{ij}=Tr(\rho\hat{P}^O_{1i}\hat{P}^O_{2j}) \label {AA22}
\end {equation}
where $\hat{P}^O$ are projectors on corresponding $O$ substates.
Corresponding subspace $L_V=V^{O1}\bigotimes V^{O2}$.

If one regards the statistical results for quantum ensembles
then statistics in $L_V$ subspace corresponds
to  $|a_j(t)|^2$ the probabilities of particular $O$
observation. Note  that their meaning differs from $O'$ 
representaion where they can't be regarded as probabilities
and can be regarded only like some weights.
Note that this restricted or partial states $R_O$ gives naturally
the values of outcome  probabilities $|a_i|^2$ which is quite difficult
to obtain in other theories explaining collapse like
Many Worlds Interpretations $\cite {Busch}$.

\section {Collapse and Quantum Memory Eraser}

To discuss measurement dynamics in our formalism for more
subtle  situations  let's consider several 
$gedankenexperiments$ for different
selfmeasurement effects : \\
1) 'Undoing' the measurement. Such experiment was discussed by Vaidman
$\cite {Vaid}$  and Detsch $\cite {De}$ for many worlds interpretation  
 but we'll regard its slightly different version. Consider
S state (\ref {AAB}) measurement by $O$ 
 resulting in the final state (\ref {AA2}). 
This S measurement  can be undone
or reversed with the help of auxiliary devices - mirrors, etc.,
which reflects $S$ back in $O$ direction and make them reinteract. 
It means that final state $\Psi_{MS}$ obtained at time $t_1$  at the
later time $t_2$ transformed backward to MS initial state $\Psi^0_{MS}$.
In any realistic layout to restore state (\ref {AAB}) is practically
impossible but to get the arbitrary S-$O$ factorized state is
more simple problem and that's enough for our considerations.
Despite that under  realistic conditions  the decoherence processes 
make this reversing immensely difficult it doesn't contradict to
any physical laws.

 If we consider this experiment in standard QM 
  we come to  some strange  conclusions. 
When memorization finished at $t_1$ in each event MS collapsed to some
 arbitrary state $|s_i\rangle|O_i\rangle$. Then at $t_2$  $O$ undergoes
the  external reversing influence, in particular it can be the
second collision with S during reversing experiment and its
state changes again and such
rescattering leads to a new state correlated with $|s_i\rangle$ :
$$
      |s_i\rangle|O_i\rangle \rightarrow |s'_i\rangle|O_o\rangle
$$
It means that $O$ memory erased and he  forgets measured
 $Q$ value $q_{i}$,
 but if he measure S state again
he would restore the same $q_i$ value.
Its statistical state is
$$
   \rho'_m=|O_0\rangle \langle O_0| \sum |a_i|^2|s'_i\rangle \langle s'_i|
$$
 But this S final state differs from
MS state (\ref {AAB}) predicted from MS linear evolution observed by  $O'$ and
in principle this difference can be tested on S state  
without $O$ measurement. 
In our doublet formalism it's necessary also to describe subjective 
event-state component $V^O$ which
 after measurement  becomes some random $V^O_j$.
But after reversing  independently of $j$ it
returns to initial value $V^O_0$ , according to evolution ansatz
 described in previous chapter.
If such description of this experiment is correct, as we can
believe because its results coincides with Schrodinger evolution in $O'$  
 it follows that after $q_i$ value erased from $O$ memory
 it lost unrestorably also
for any  other possible observer. If after that $O$ would measure Q again
 obtained new value $q_j$ will have no correlation with $q_i$.

Of course one should remember that  existing for finite time intermediate
$O$ states are in fact virtual states and differ from really stable 
states  used here, but for macroscopic time
 intervals this difference becomes
very small and probably can be neglected.

The analogy of 'undoing' with  quantum eraser experiment is straightforward :
there the photons polarization carry the information 
which can be erased and so change the system state $\cite {Scu}$.   
The analogous experiment with information memorization by 
some massive objects like molecules will be important test of
collapse models.

Note that observer $O'$ can perform on $O$ and S also the direct measurement
of interference terms for ($\ref {AA2}$) without reversing MS state.
Such experiment regarded for Coleman-Hepp model in  $\cite {May3}$
doesn't introduces any new features in comparison with 'Undoing'
and so we don't discuss it here.

2) After $O$ measures $Q$ value of S at $t_1$ which results in
MS state (\ref {AA2}) for $O'$, this S observable is measured 
again by observer $O'$ at $t_2>t_1$.
The interaction of $O'$ with MS results in entangled state of 
S,$O$, and $O'$ and so both observers acquire  some information 
about S state. $O'$, MS state relative to next observer $O^2$ is:
\begin {equation}
   \Psi'_{MS}=|a_1 |s_1\rangle|O_1\rangle|O'_1\rangle+
a_2|s_2\rangle|O_2\rangle|O'_2\rangle
                                   \label {AAX}
\end {equation}   
Note that now  there is interference term operator $B'$ on $O'+$ MS
  which is unobservable for $O'$.
 The experiments of such kind were discussed  frequently, due to
its relation to EPR-Bohm correlations, but here we regard in detail
 its timing sequence.
In our formalism at $t_1<t<t_2$ observer $O$ already have the information
that Q value is some  $q_i$, reflected by $V^O=|O_i\rangle\langle O_i|$.
 In the same time Q value stays
 uncertain and objectively nonexisting
 for $O'$, because relative to him MS state is (\ref {AA2})
,  which isn't Q eigenvalue $\cite {Busch}$.
 When measurement by $O'$ 
 finished the obtained Q value coincides with $q_i$, but
it don't contradicts to the previous assumption 
 that for $O'$ before $t_2$ it was principally uncertain.
The reason is that in between $O'$ interacts with S and this interaction
transfer S information to $O'$ and makes Q value definite for him. 
To check that Q value coincides for $O'$ and $O$, $O'$ can perform 
measurement both Q and $Q_O$ which is described by (\ref {AA22}).  
Together this experiments  supposedly  demonstrate the subjective character
of collapse, which happens only after interaction of S with
particular observer occurs. It differs from  standard QM picture
of objective collapse which occurs  for all observers simultaneously
independently of their participation in the measurement.
This results contradict to 
 first intuitive impression  that if Q had 
some definite value relative to 
$O$ then its objectively exists also for $O'$ and any other
observer.
But it's erroneous conclusion because at that time MS state relative to
$O'$  is pure state $\Psi_{MS}$ of (\ref {AA2})  which isn't $Q$
eigenstate. To demonstrate it experimentally $O'$ can measure $\hat{B}$ on MS
which don't commute with Q and  for which $\Psi_{MS}$ is eigenstate.
Alternatively  $O'$ can perform 'undoing' on MS and Q value
 known to $O$ will be erased unrestorably, which is impossible if
Q value objectively existed for $O'$. 

Let's discuss   why $O, O'$ MS descriptions can be compatible
without contradictions. After MS selfmesurement finished in a state
(\ref {AA2})  it describes $O'$ information
about S,$O$ states and in particular Q uncertainty. 
Even if $O$ have  definite information about Q value, as our doublet formalism
assume, until any signal will be send by $O$ to $O'$ his information
don't change and described by $\Psi_{MS}$ of (\ref {AA2}).
 S measurement by $O'$ discussed above corresponds to such signal
and after it $O'$ acquire new information expressed by $V'^O$ value.

Relativistic analysis of EPR-Bohm pairs measurement  also indicates
subjective character of state vector and its collapse $\cite {Aha2}$.
 It's shown
that state vector can be defined only on space-like hypersurfaces
which are noncovariant for different observers.
This results supports nonequivalence of different observers
assumed in Relational QM and our formalism, for which EPR-Bohm correlations
seems to deserve detailed study. 

 In case of general S-$O$ interaction 
 consisting of several effective  rescatterings alike in 
'undoing', one should
calculate $P_i(t)$ each time and define  $V^O$ anew.
 Note that we must be really interested only in final $O$ state, because
in this model $O$ has no memory about intermediate  states and no
dynamical dependence on them.

In doublet formalism $O$ percepts only $O_j$ component of 
complete state vector $\Psi_{MS}$ and it's interesting to
speculate why other its components aren't observed.
Our tempting explanation is prompted by Breuer theorem, but in doublet
formalism we should reconsider it for
   probabilistic situation. It was shown that $O$ 
selfmeasurement is always noncomplete and it's possible to assume
that  in given  event $O$ can percept only  part  $O_j$ of
his effective physical state. Note that in the same time
$O'$ can perform $B$ measurement which demonstrates the existence
of other $O$ components. 

Our doublet formalism can be  interpreted if to concede that
S initial state vector describes $Q$ fundamental uncertainty  for $O$ i.e.
limited amount of selected information $I_Q$.
 When S-$O$ start to interact
corresponding $O$ internal DFs excited and its internal state correlated  
with S. It's tempting to assume that for $O$ internal states any
uncertainty is excluded - i.e. $O$ knows his own state
due to continuous interactions inside $O$ and initially uncertain Q
 percepted as random but certain value $Q_O$.
$O$ is the 'last ring' of measurement chain and it can have such
singularity, because it related to $O$ uncertainty relative to himself
and don't contradicts to QM dynamics. 
 It reminds
Von Neuman and London - Bauer collapse theories, but our formalism
agrees with applicability of Schrodinger equation to MS
for external observer $O'$.

\section {Discussion}

In this paper the simple measurement model studied
 which accounts observer (IGUS) information
processing and memorization. Real IGUSes are very complicated systems 
with many DFs, but the main quantum effects like superpositions or
decoherence are the same for large and small systems and 
can be studied with the simple models.

Summing up our results we notice that by itself observer inclusion
into QM measurement chain doesn't lead to collapse explanation.
If in addition QM states manifold changed to dual structure
it results in consistent collapse description, which 
 permit to change collapse postulate   modifying it 
into more consistent and sensible form.
Its most important feature is the absence of any special collapse
dynamics  for external observer $O'$ for which MS evolves
according to Schrodinger equation.

So it seems that to avoid completely even weak  or subjective collapse
 it's necessary to
 reconsider the  quantum theory foundations, not only its measurement
part. For this purpose we develop in forcoming paper
Information Causality Interpretation (ICI) which explain the appearance
of doublet event-state formalism. 

Our doublet formalism demonstrates that probabilistic evolution
 is generic and unavoidable for QM and without it QM can't
acquire any sensible observational realization. Wave-particle dualism
was always regarded as characteristic QM  feature, but in our formalism
it has straightforward  description. 

Rovelli assumed that in QM all physical variables can have definite values
only relative to some observer or RF. Correspondingly in Relational QM
state collapse is subjective effect occuring in the interaction
of $O$ with measured state or with the signal send by other observer.
It seems that doublet formalism can describe
this main effects predicted phenomenologically by Relational QM.

It's widely accepted now that decoherence effects are very important
in measurement dynamics $\cite {Zur,Gui}$. But the frequent claim
that collapse phenomena  can be completely explained in its framework
was shown to be incorrect at least for simple models $\cite {Desp}$.
But in our model of subjective or weak collapse some kind of decoherence
is also present in the form of self-decoherence when S departs from
$O$ after interaction and additional $O$-environment decohering 
interaction only will amplify this effects.
Our approach to collapse is close to the decoherence attitude,
where also any additional collapse dynamics don't exist. The main difference
is that  we suppose
that collapse isn't objective phenomena
and has relational or subjective character and observed only by
observer inside decohering system.

The situation with the measurement problem for two quantum observers
has much in common with Quantum reference frames introduced by
 Aharonov  $\cite {Aha3}$.
Note that our formalism is principally different from hidden parameters
Theories where this stochastic parameters influence Quantum state
dynamics. In our model  $V^O$ internal parameter $j$ is on the opposite
is controlled by evolution equation for quantum state.

Our formalism deserves detailed comparison with formalisms of different MWI 
variants, due to their analogy  - both are theories without
dynamical collapse  $\cite {Busch}$.
In Everett+brain QM interpretations eq. ($\ref {AA2}$)
describes so called observer $O$ splitting identified with state collapse
 $\cite {Whi}$.  In this theory it's  assumed that each
$O$ branch describes the different reality and the state collapse is
phenomenological property of human consciousness. Obviously this
approach has
some common points with our models which deserve further analysis. 
In general all our experimental conclusions are based on human
subjective perception. Assuming the computer-brain perception analogy
in fact means that human signal perception also defined by $\bar{Q}_O$
values. Despite that this analogy looks quite reasonable we can't
give any proof of it.  In our model
in fact  the state collapse have subjective character and 
occurs initially only for single observer $O$, but as was shown
by Rovelli it doesn't results in any contradictions $\cite {Rov}$. 
If it's sensible to discuss any world partition prompted by QM results
it seems to be the division between subject - observer $O$
 which collect information
about surrounding world  and  this world objects  which can include other
observer $O'$.

\section* {Appendix}

The simple model of observer with many DFs
  is modified Coleman-Hepp (CH) model  which
 used  often for QM paradoxes discussion $\cite {Hep}$.
CH model considers  fermion $S^0$ spin z-projection measurement via
interaction with $N$ spin-half atoms $A_i$ linear chain -
1-dimensional crystal detector  D. $A_i$
atoms are regularly localized at the distance $r_0$ by the effective potential
 $U_i(x_i)$. $S^0$ initial 
state $\psi^0_0=\varphi(x,t_0)(a_1|u_0\rangle+a_2|d_0\rangle)$ 
where $u,d$ are up,down spin states and $\varphi(x,t_0)$ is localized $S^0$
 wave packet spreading along  D spin chain. 
For the comparison  the measurement of corresponding
 mixed state $\rho^0_m$ with   weights $|a_{1,2}|^2$ will be regarded.
 $S^0 - D$ interaction Hamiltonian  is:
\begin {equation}
   H_I=(1-\sigma^0_z)\sum^N_{i=1} V(x-x_i) \sigma^i_x  \label {A0}
\end {equation}
where $V$ is $S^0-A_i$ interaction potential.
 For suitable  model parameters and for
 D initial polarized state  $\psi^D_+=\prod|u_i\rangle$ one obtains that
 if $S^0$ initial spin state is $|u_0\rangle$ this D state conserved
 after $S^0$ passed over the  chain, 
but for initial state $|d_0\rangle$ 
D state transformed into $\psi^D_- =\prod|d_i\rangle$. 
Thus for  finite $N$ at $t>t_1$ for $S^0-D$ final state
\begin {equation}
 \psi_f(t)=\psi_1(t)+\psi_2(t)=
\varphi(x,t)(a_1|u_0\rangle\psi^D_++a_2(-i)^N |d_0\rangle\psi^D_- )
                                  \label {A2}
\end {equation}
we get macroscopically different values of
D pointer which described by the polarization operator  :
 $\mu_z=\frac{1}{N}\sum \sigma^i_z$ acting in $h_D$ subspace.
It gives estimate $\bar{\mu}_z=\bar{\sigma}^0_z$ and 
$\bar{\Delta}_Q=0$, so this is strict exact
measurement.  
Despite, it doesn't mean $S^0$ state collapse because $S^0 - D$
interference terms (IT) operator:
\begin {equation}
B=\sigma^0_x B_I=\sigma^0_x\prod_{i=1}^N\sigma^i_y   \label {A2A}
\end {equation}
describing spin-flips of all $A_i$ and $S^0$ spins.
In principle $B$ also can be measured by
observer $O$ and discriminate $S^0 - D$ mixed and pure states.
Its expectation value $\bar{B}=.5(a_1^*a_2+a_1a_2^*)$
 for $S^0 - D$  final state $\psi_f$ differs from
$\bar{B}=0$ for $S^0$ mixed  state $\rho^0_m$ $\cite {Bel}$.
For the convenience we exclude from consideration
 $a_1, a_2$ values such that $\bar{B}=0$,
which doesn't influence on our final results.
 Note that $S^0$ IT can be measured separately, but
 only before $S^0-D$ interaction starts, after it only their joint IT operator
have sense.
 $\mu_z,B$ don't commute and can't be measured
simultaneously:
\begin {equation}
  [\mu_z,B]=\frac{i\sigma^0_x}{N}
\sum_{i=1}^N \sigma^i_x\prod^N_{i\neq j}\sigma^j_y
            \label {A3}
\end {equation}
It's easy to propose how to measure collective (additive) operator $\mu_z$,
 but also 
$B$ values can be destructively measured  decomposing D into atoms
and sending $A_i$ one by one and also $S^0$ into Stern-Gerlach magnet.
 Then measuring
$A_i$ amount in each channel and their correlations by some other detector
D$'$ one obtains information on $B$ value from it.
Standard QM don't regard any special features of
 destructive measurements assuming that any hermitian operator
is observable and can be measured by one way  or another.

\end{document}